\documentstyle[12pt]{article}
\begin{document}
\pagestyle{plain}
\setcounter{page}{1}
\newcounter{bean}
\baselineskip16pt


\begin{titlepage}
\begin{flushright}
PUPT-1671\\
hep-th/9612051
\end{flushright}

\vspace{20 mm}

\begin{center}
{\huge Fixed scalar greybody factors }
\vspace{5mm}
{\huge in five and four dimensions}
\end{center}
\vspace{10 mm}
\begin{center}
{\large  
Igor R.~Klebanov and Michael Krasnitz\\
}
\vspace{3mm}
Joseph Henry Laboratories\\
Princeton University\\
Princeton, New Jersey 08544
\end{center}
\vspace{2cm}
\begin{center}
{\large Abstract}
\end{center}
\noindent
We perform the classical gravity calculations of 
the  fixed scalar
absorption cross-sections by $D=5$ black holes with
three charges and by $D=4$ black holes with four charges. 
We obtain analytic results for the cases where the energy and
the left and right moving temperatures are sufficiently low
but have arbitrary ratios.
In $D=5$ the greybody factor is in perfect agreement with the
recent calculation performed in the context of the effective string
model for black holes.
In $D=4$ the formula for the greybody factor in terms of the energy and
the temperatures differs from that in $D=5$ only by the overall normalization.
This suggests that the fixed scalar coupling to the
effective string in $D=4$ is identical to that in $D=5$. 

\vspace{2cm}
\begin{flushleft}
December 1996

\end{flushleft}
\end{titlepage}


\newpage
\renewcommand{\baselinestretch}{1.1} 


\renewcommand{\epsilon}{\varepsilon}
\def\fixit#1{}
\def\comment#1{}
\def\equno#1{(\ref{#1})}
\def\equnos#1{(#1)}
\def\sectno#1{section~\ref{#1}}
\def\figno#1{Fig.~(\ref{#1})}
\def\D#1#2{{\partial #1 \over \partial #2}}
\def\df#1#2{{\displaystyle{#1 \over #2}}}
\def\tf#1#2{{\textstyle{#1 \over #2}}}
\def\d{{\rm d}}
\def\e{{\rm e}}
\def\i{{\rm i}}
\def\Leff{L_{\rm eff}}


\section{The $D=5$ black holes}
\subsection{Introduction}
\label{Intro}

Recently remarkable progress has been achieved in 
describing $D=5$ black holes with three different $U(1)$ charges 
\cite{sv,arkady,cm,hms} in 
the string theory context. In the extremal limit these black holes preserve
$1/8$ of the original supersymmetry and also have a finite horizon area.
They may be embedded into string theory
using intersecting D-branes, and the resulting entropy
of string states agrees with the 
Bekenstein-Hawking entropy \cite{sv,cm,hms}. 
It is believed that their low-energy dynamics
is described by small fluctuations of a long intersection string
\cite{ms}. 
The model involves $n_1$ 1-branes 
marginally bound to $n_5$ 5-branes, with some
longitudinal momentum along the 1-branes carried by left moving
open strings. In the near-extremal case, right movers are also
present, so that a left moving and a right moving open string may
collide to produce an outgoing closed string \cite{cm,hk}. 
The inverse of this process, which gives the
leading order contribution to the absorption of closed strings, was
also found to be in agreement with the semiclassical gravity, up to
an overall normalization \cite{dmw}.   
Das and Mathur \cite{dm} subsequently normalized the leading
emission and absorption rates, both in semiclassical gravity and
in the D-brane picture, and found perfect agreement.  The specific picture
used in \cite{dm} follows that suggested in \cite{dm1,ms}:  the low-energy
dynamics of the D-brane configuration is captured by a single string with
winding number $n_1 n_5$ which is free to vibrate only within the 5-brane
hyperplane.  The calculation 
of emission and absorption was generalized to charged particles in
\cite{us}. Furthermore, 
Maldacena and Strominger \cite{mast} 
showed that the agreement between the effective string model and the
general relativity continues to hold when the energy, $\omega$, and the
left and right moving temperatures,
$T_L$ and $T_R$, are all comparable. In terms of the
four radii of the black hole, the parameter region considered in 
\cite{mast} is
\begin{equation} \label{dilute}
r_0, r_K \ll r_1, r_5 \ .
\end{equation}
The greybody factor in this regime has the dependence on
$\omega$,
$T_L$ and $T_R$ which provides strong evidence in favor of the
effective string model of $D=5$ black holes \cite{mast}.

Even more intricate evidence in favor of this model was recently provided
by the calculations of the fixed scalar greybody factors \cite{cgkt}.
The fixed scalars \cite{fs}
are the special massless fields whose values on
the horizon of an extremal black hole are fixed by the $U(1)$ charges.
This fixing translates into a suppression of the absorption and emission at
low energies compared to the ordinary massless scalars \cite{kr}.
In  \cite{cgkt} it was shown that the fixed scalars couple 
to the effective string
differently than the ordinary scalars. 
A specific fixed scalar $\nu$, related to
the volume of $T^4$ around which the 5-branes are wrapped, 
was found to couple to the world sheet as
\begin{equation}
{1\over 4 T_{\rm eff}}\int d^2\sigma T_{++} T_{--} \nu
\label{coupling}\end{equation}
where $T_{\rm eff}$ is the string tension, while
$T_{++}$ and $T_{--}$ are the left and right 
moving components of
the stress-energy tensor on the effective string. 
The effective string
calculation of the absorption cross-section, which utilized the
methods of thermal field theory, 
yielded a simple analytic formula \cite{cgkt},
\begin{equation}
\label{greybody}
\sigma_{\rm abs} = {\kappa_5^2 L_{\rm eff}\over (32 \pi T_{\rm eff})^2}
(\omega^2+16\pi^2T_L^2)(\omega^2+16\pi^2T_R^2)\ \omega{
e^{\omega\over T_H}-1 
\over (e^{\omega\over 2T_L}-1)(e^{\omega\over 2T_R}-1)} 
\end{equation}
where $\kappa_5$ is the 
$D=5$ gravitational constant, and $L_{\rm eff}$ is the length
of the effective string, which is related to the radii by
\begin{equation}
\label{length}
\kappa_5^2 L_{\rm eff}= 4\pi^3 r_1^2 r_5^2
\ .\end{equation}
In \cite{cgkt} the greybody factor (\ref{greybody}) was shown to agree with
the general relativity
absorption calculation carried out in the extremal ($T_R =0$) and
the near-extremal ($T_R \ll T_L$) cases. 
For technical reasons, these calculations
were carried out for $r_1=r_5=R$, and the effective string tension
required for the agreement was found to be
\begin{equation}
\label{tension}
 T_{\rm eff} = {1\over 2\pi R^2}\ . 
\end{equation}
Since $R^2$ grows as the number of D-branes,
this formula confirms the idea that the effective string
is fractionated \cite{juan,halyo}. 

In this paper we carry out a more general classical absorption calculation,
and show its complete agreement with (\ref{greybody}). We find it
possible to reduce the fixed scalar equation in the inner region
(near the horizon) to the hypergeometric equation. After a matching of
approximate solutions, the result (\ref{greybody}) follows from the
well-known asymptotics of the hypergeometric functions.

\subsection{The classical absorption calculation}

The Einstein metric for the five-dimensional black hole
is \cite{arkady,cm,cy1,hms} 
$$ ds^2 = -f^{-2/3}hdt^2 + 
f^{1/3}(h^{-1}dr^2 + r^2d\Omega _3 ^2)\ , $$
where 
$$ h(r) = 1-{r_0^2\over r^2}, 
\qquad f(r) = (1 + {r_K^2\over r^2})(1 + {r_1^2\over r^2})(1 + {r_5^2\over r^2})\ . $$
$r_0$ is the non-extremality parameter of the black hole, while 
$r_K, r_1, r_5$ are related to the 
other charges. One also introduces the hyperbolic angle $\sigma$ defined by
$$ r_K = r_0 \sinh \sigma $$
The left and right temperatures are \cite{hms,mast} 
$$ T_L = {r_0 e^ \sigma \over 2 \pi r_1r_5} , 
\qquad T_R = {r_0 e^ {- \sigma} \over 2 \pi r_1r_5}, $$
and the Hawking temperature is their harmonic average:
$$ {2\over T_H} = {1\over T_L} + {1\over T_R}\ . $$
We will confine ourselves to the dilute gas region (\ref{dilute}) and
consider sufficiently low frequencies,
$$ \omega r_i \ll 1 \ .$$ 
This is the parameter
region decribed by the effective string model. 

The equation describing the propagation of the fixed scalar $\nu$
near a black hole with
$r_1 = r_5 = R $ (only in this case does it seem possible to obtain a 
simple equation) was derived in \cite{cgkt}, 
\begin{equation}
\left [(hr^3{d \over dr})^2 + (r^2+R^2)^2(r^2+r_K^2)\omega ^2 - 
{8r^4 R^4 \over (r^2 +R^2)^2}h \right ] \nu(r) = 0 \ .
\label {full5}
\end{equation}
We solve (\ref{full5})
 by the matching technique as in \cite{dm,mast,cgkt}. 
Namely, we divide space into three regions, 
in each of which the equation simplifies and can be solved explicitly: 

I. The Near region: $ r \ll R $.

II. The Middle region: $  r_0 \ll r \ll  1/\omega $.

III. The Far region: $ r \gg R $.

\noindent
Note that, since $  r_0 \ll R \ll  1/\omega $, the
middle region overlaps each of the other two.

Now we show how (\ref{full5}) simplifies in each of the three
regions.
In regions II and III we may approximate $h(r) = 1 $. 
In region III the potential 
is negligible compared to the frequency term, and the equation reduces to 
$$\left [r^{-3}{d\over dr}r^3{d\over dr} + 
\omega ^2 \right ]\nu_{III} = 0\ . $$
The solution may be written as 
$$ \nu_{III} = \alpha {J_1(\omega r)\over \omega r} + 
\beta {N_1(\omega r)\over \omega r} $$
where $J_1, N_1$ are correspondingly the Bessel and Neumann functions.

In region II the frequency term may be 
neglected, and we have the approximate equation
$$\left [(r^3{d\over dr})^2 - 
8{R^4\over H^2} \right ]\nu_{II} = 0 $$
with $$ H(r) = 1 + {R^2\over r^2}\ . $$
The solution has two undetermined parameters,
$$\nu_{II} = {A\over H(r)} + BH^2(r)\ . $$

In region I (\ref{full5}) becomes
$$ \left [(hr^3{d\over dr})^2 + 
R^4(r^2+r_K^2)\omega ^2 - 8r^4h \right ]\nu_I = 0\ . $$
In terms of the variable $ z = 1 - {r_0^2\over r^2} $ we get
\begin{equation}\label{regionone}
\left [(z{d\over dz})^2 + D + 
{C\over 1-z} - {2z\over (1-z)^2}\right ]\nu_I = 0 
\ .\end{equation}
where
\begin{equation}
D = {\omega ^2 (T_L-T_R)^2\over 64\pi ^2 (T_LT_R)^2} \ ,
\qquad  C={\omega ^2\over 16\pi ^2T_LT_R}
\ .\label{CD}
\end{equation}
In the immediate vicinity of the horizon, 
we have $ z \ll 1 $, and the equation reduces to
$$\left  [(z{d\over dz})^2 + (C+D) \right ] \nu_I = 0\ . $$
The incoming solution at the horizon is 
$$\nu_I (z\rightarrow 0)= e^{-i\sqrt{C+D}\log z} =z^{-i(a+b)/2}
\ ,
$$
where following \cite{mast} we defined
\begin{equation}
a = {\omega\over 4\pi T_L}\ , 
\qquad b = {\omega\over 4\pi T_R}\ .
\label{ab}
\end{equation} 
To solve the full region I equation, (\ref{regionone}),
we will define a new function $F$ by
$$\nu_I (z)= z^{-i(a+b)/2} (1-z)^{-1} F(z)\ .$$ 
Substituting this into (\ref{regionone}), we see that
that $F$ satisfies a hypergeometric equation. 
\begin{equation}
z(1-z){d^2 F\over dz^2} + [ (1- ia - ib) + (1 + ia + ib)z]{dF\over dz} - 
(1+ia)(1+ib)F = 0\ .
\label{hyper}
\end{equation}
In order to get an incoming solution near the horizon, 
we choose the boundary condition $F(0)=1$.
The solution is the hypergeometric function $F(-1-ia,-1-ib,1-ia-ib;z)$.

To match this with the region II solution we take the limit $z\rightarrow 1$. 
Defining $v=1-z$, we expand around $v=0$:
$$F(-1-ia,-1-ib,1-ia-ib;1-v) = E + Gv + O(v^2)$$ 
where
$$ E = {2\Gamma(1-ia-ib)\over \Gamma(2-ia)\Gamma(2-ib)} $$
and $G$ is a constant of the same order as $E$ whose value 
will not be important to us.
As a consequence,
$$\nu_I (v\rightarrow 0) = Ev^{-1} + G+O(v) = 
E{r^2\over r_0^2}+G+ O(v) $$
We match this to the small $r$ behavior of the
middle region solution,
$$ \nu_{II} = A{r^2\over R^2} + B{R^4\over r^4}\ , $$
where we have kept the dominant term in the expansion
of $H(r)$, by requiring that $\nu$ 
and its derivative are continuous at $r= r_m$.
The matching point $r_m$ satisfies
$$r_0 \ll r_m \ll R\ . $$
Solving the two matching equations, we obtain
$$ A = E{R^2\over r_0^2}+G {2 R^2\over 3 r_m^2}
\ ,\qquad B = G{r_m^4\over 3 R^4}\ . $$
We see that $ B \ll A $.
In fact, throughout
region II, $BH^2$ may be neglected relative to
$A H^{-1}$. 
We can also neglect the second term in
$A$ relative to the first one and use $A= E{R^2\over r_0^2}$. 

We now match regions II and III. As $\omega r\rightarrow 0$, we have
$$\nu_{III}\rightarrow {\alpha\over 2} - {2\beta\over \pi\omega^2r^2} $$
and, as ${R\over r}\rightarrow 0 $, we have
$$\nu_{II}\rightarrow A(1-{R^2\over r^2}) \ .$$
Matching, we obtain
$$ \alpha = 2A \ ,\qquad  \beta = {\pi A \omega^2R^2\over 2}\ . $$
We see that 
$$ {\beta\over \alpha} \sim (\omega R)^2 \ll 1 $$
so that we may neglect $\beta $ in what follows.
We obtain
$$ \alpha = 2A = 2E{R^2\over r_0^2}\ . $$ 

The absorption probability is the ratio 
of the incoming fluxes at the horizon and at infinity. 
In this case the flux is given by
$$ {\it F}={1\over 2i}(\nu^* h r^3 {d\over dr}\nu - c.c.)$$
and so we get 
$$P_{\rm abs}= { {\it F}_{\rm horizon}
\over {\it F}_\infty}
= {2\pi\over |\alpha|^2}R^2\sqrt{r_0^2+r_K^2}\omega ^3\ . $$
By the Optical Theorem in $D=5$, the absorption cross-section is \cite{dm}
$$ \sigma_{\rm abs} = {4\pi \over \omega ^3} P_{\rm abs}
\ . $$
Using the identities
$$ \Gamma (z+1) = z \Gamma (z)\  ,
\qquad |\Gamma (1-ia)|^2 = {\pi a\over \sinh \pi a}\ , $$
we obtain the following value for the absorption cross-section,
\begin{equation}
\sigma_{\rm abs} = {\pi^3R^8\over 64}(\omega^2+16\pi^2T_L^2)
(\omega^2+16\pi^2T_R^2)\omega{e^{\omega\over T_H}-1 \over 
(e^{\omega\over 2T_L}-1)(e^{\omega\over 2T_R}-1)} \ .
\label{sigma5}
\end{equation}
Using (\ref{length}) 
and (\ref{tension}) we find that this is in precise agreement with
the absorption cross-section predicted by the effective string model,
(\ref{greybody}).

\section{The D=4 Case}
\subsection{Introduction}

The $D=4$ black holes with four $U(1)$ charges \cite{cy,ct}
have many features in common with
the $D=5$ black holes considered above.
An effective string model for such $D=4$ black holes is motivated by
their embedding into M-theory \cite{at,kt}.
A specific
configuration useful for explaining the Bekenstein-Hawking entropy is
the $5\bot 5\bot 5$ intersection \cite{kt}: there are $n_1$ 5-branes
in the $(12345)$ hyperplane, $n_2$ 5-branes in the $(12367)$
hyperplane, and $n_3$ 5-branes in the $(14567)$ hyperplane. One also
introduces a left moving momentum along the intersection
string (in the $\hat 1$ direction). If the length of this direction is
$L_1$, then the momentum is quantized as $2\pi n_K/L_1$, 
so that $n_K$ plays the role of the fourth $U(1)$
charge.  Upon compactification on $T^7$ the metric of the $5\bot 5\bot
5$ configuration
reduces to that of the $D=4$ black hole with four charges.  Just like
in the D-brane description of the $D=5$ black hole, the low-energy
excitations are signals propagating along the intersection string. In
M-theory the relevant states are likely to be small 2-branes with
three holes glued into the three different hyperplanes \cite{kt}. As a
result, the effective length of the intersection string is
$\Leff= n_1 n_2 n_3 L_1$. 
This fact, together with the assumption that these
modes carry central charge $c=6$, is enough to reproduce the
extremal 
Bekenstein-Hawking entropy, $S=2\pi\sqrt{n_1 n_2 n_3 n_K}$ \cite{kt}.
In \cite{us} it was shown that this ``multiply-wound
string'' model of the four-charge $D=4$ black hole  correctly
reproduces the Hawking radiation of both neutral and Kaluza-Klein
charged scalars. Furthermore, the
ordinary scalar greybody factor agrees with that in
the effective string picture 
of the $D=4$ black holes \cite{gk}. 
The parameter region considered in \cite{gk} is:
\begin{equation}
r_0, r_K\ll r_1, r_2, r_3 \ . 
\label{newchoice}\end{equation}
In this section we calculate the fixed scalar greybody 
factor in this region and show that, up to the overall
normalization, it is identical to that in the
$D=5$ case. 
We conclude by discussing possible implications of
this equivalence for the 
coupling of the fixed scalars to the effective string.

\section{The semiclassical gravity analysis}

The metric of the $D=4$ black hole 
with four charges is \cite{cy,ct,ct1}
$$ ds^2 = -f^{-1/2}hdt^2 + f^{1/2}(h^{-1}dr^2 + r^2d\Omega _2^2) $$
where
$$ h(r) = 1 - {r_0\over r}
\ ,\qquad f(r) = (1+{r_K\over r})
(1 + {r_1\over r})(1+{r_2\over r})(1+{r_3\over r}) $$
Now we define $\sigma$ by
$$ r_K = r_0 (\sinh \sigma)^2\ . $$
The left and right temperatures are \cite{gk}
$$ T_L = {1\over 4\pi}\sqrt{{r_0\over r_1r_2r_3}}e^{\sigma}
\ ,\qquad T_R = {1\over 4\pi}\sqrt{{r_0\over r_1r_2r_3}}e^{-\sigma}
\ ,
$$
and the Hawking temperature is their harmonic average.
We will work in the dilute gas region (\ref{newchoice})
and assume the low energy condition,
$$ \omega r_i \ll 1\ . $$

The equations governing the fixed scalar propagation 
for this range of parameters were
studied in \cite{cgkt}. 
To get a tractable equation for the fixed scalar 
fluctuations, we may take
three large charges to be equal: 
$$ r_1 = r_2 = r_3 = R $$
Then we have the following equation for the fixed scalar $\nu$,
\begin{equation}
\left [(hr^2{d\over dr})^2 + \omega^2r^4(1+{r_K\over r})(1+{R\over r})^3-
2hR^2(1+{R\over r})^{-2} \right ]\nu = 0 \ .
\label{full4}
\end{equation}
As before, we solve this by considering the near, 
middle and far regions defined in the same way as for the 
$D=5$ case. 

In the far region we get the solution
$$\nu_{III} = \alpha {\sin(\omega r)\over \omega r} + 
\beta {\cos(\omega r)\over \omega r}\ , $$
while in the middle region the solution is
$$\nu_{II} = A \left (1+{R\over r} \right )^{-1} + 
B \left (1+{R\over r} \right )^2\ . $$
As before we find that
$$ \beta \ll \alpha\ ,\qquad B \ll A\ , $$
and this time 
$$ A = \alpha \ .$$
In the near region we get precisely 
the same equation as for $D=5$,
(\ref{regionone}), with $a,b,C,D$ defined 
through $T_L, T_R$ as in (\ref{CD}) and  (\ref{ab}),
and $z=1- {r_0\over r}$. 
Performing the matching in the same way as before, we get
$$ \alpha = {2R\over r_0} {\Gamma(1-ia-ib)\over 
\Gamma(2-ia)\Gamma(2-ib)}\ . $$

It is again convenient to
find the absorption probability with the method of fluxes. 
Now the radial flux is given by
$$ {\it F} = {1\over 2i}(\nu^*hr^2{d\over dr}\nu - c.c.)\ . $$
The absorption probability is calculated to be
$$P_{\rm abs}= {r_0\omega^2\over \pi T_H|\alpha|^2}\ . $$
By the Optical Theorem,
$$ \sigma_{\rm abs} = {\pi\over \omega^2} P_{\rm abs} $$
and we find
\begin{equation}
\sigma_{\rm abs}={\pi^2R^7}(\omega^2+16\pi^2T_L^2)(\omega^2+16\pi^2T_R^2)
\omega{e^{\omega\over T_H}-1 
\over (e^{\omega\over 2T_L}-1)(e^{\omega\over 2T_R}-1)}\ . 
\label{sigma4}
\end{equation}
Thus, up to overall normalization, the fixed scalar greybody factor in
$D=4$ is identical to that found in $D=5$. 

While we do not understand the effective string for the $D=4$ black holes
as well as in the $D=5$ case, it is tempting to conjecture that,
up to the overall normalization, the coupling of the $D=4$ fixed 
scalar is given by (\ref{coupling}). Then the effective string
result for the cross-section is proportional to (\ref{greybody}) with
$\kappa_5$ replaced by $\kappa_4$. Comparing this to (\ref{sigma4}) and
using the relation \cite{gk}
\begin{equation}
\kappa_4^2 L_{\rm eff}=  16 \pi^2 r_1 r_2 r_3= 16 \pi^2 R^3
\label{chargerel}\end{equation}
we find that
the effective string tension scales as 
$$ T_{\rm eff} \sim {1\over R^2}\ . $$
If we denote by $n$ the number of 5-branes in each of the three orientations,
we see that $T_{\rm eff}\sim n^{-2}$ (because the black hole
radius scales as $R\sim n$). Interestingly, 
this agrees with the scaling of the
string tension necessary to explain the entropy of the 
near-extremal $5\bot 5$
configuration \cite{ami,kt1}. \footnote{
A different interpretation of the rescaling of the string tension
was given in \cite{lw,ct,halyo,eh}.}
As was argued in \cite{ami}, this is precisely the 
string that arises at the triple intersection of M5-branes.

\section{Concluding Remarks}

This paper contributes to a growing list of similarities between
$D=5$ black holes with three charges and $D=4$ black holes with
four charges. In both cases, the 
microscopic interpretations of the entropy
are provided by effective strings whose physical modes carry central charge
$c=6$. Furthermore, the classical gravity calculations of the
greybody factors for ordinary 
massless scalars reveal that they have identical 
coupling to the effective string \cite{gk}. 
Perhaps this was not a surprise since 
this coupling was given by the minimal two-derivative term.
Now we find that even the peculiar fixed scalars have essentially
identical greybody factors in $D=5$ and $D=4$. In $D=5$ the greybody
factor coincides exactly with that produced by the coupling to
$T_{++} T_{--}$ on the effective string \cite{cgkt}. This is also the case 
in $D=4$, although we do not yet know of a derivation of this coupling
from an effective string action.\footnote{We are grateful to A. Tseytlin
for many discussions on this point.} 
Based on the accumulating evidence, we feel that one should be able to
construct a unified picture of the $D=5$ and $D=4$
black holes which shows that the effective strings responsible for
their microscopic degrees of freedom have a common origin.


\section*{Acknowledgements}

This article is based on a Princeton University Junior Paper by one of
us (M.K.).
We are grateful to C.G.~Callan, S.S.~Gubser and
A.A.~Tseytlin for useful discussions and comments.  The work of
I.R.~Klebanov was supported in part by DOE grant DE-FG02-91ER40671,
the NSF Presidential Young Investigator Award PHY-9157482, and the
James S.{} McDonnell Foundation grant No.{} 91-48.  


\end{document}